# Pain and Spontaneous Thought

Aaron Kucyi, PhD

*Department of Neurology and Neurological Sciences, Stanford University, Stanford, CA, USA,*

*94304*

**Abstract**

Pain is among the most salient of experiences while also, curiously, being among the most malleable. A large body of research has revealed that a multitude of explicit strategies can be used to effectively alter the attention-demanding quality of acute and chronic pains and their associated neural correlates. However, thoughts that are spontaneous, rather than actively generated, are common in daily life, and so attention to pain can often temporally fluctuate because of ongoing self-generated experiences. Classic pain theories have largely neglected to account for unconstrained fluctuations in cognition, but new studies have demonstrated the behavioral-relevance, putative neural basis, and individual variability of interactions between pain and spontaneous thoughts. In this chapter, I review behavioral studies of ongoing fluctuations in attention to pain, studies of the neural basis of spontaneous mind-wandering away from pain, and the clinical implications of this research.

**Introduction**

An overwhelming proportion of studies on pain-attention interactions involve either an intervention to explicitly manipulate cognitive state or an instruction requiring an individual to actively control or distract from pain. In naturalistic settings, however, attention to acute and chronic pain can often fluctuate because of ongoing spontaneous thoughts. Whether a person's thought contents during a tongue piercing procedure are focused on negative or positive aspects of the procedure, or are completely unrelated, may not be under their active control and may not even be directly influenced by the present sensory environment. While classic pain theories have largely neglected to account for such unconstrained fluctuations in cognition, spontaneous thoughts are a defining part of daily life— as illustrated throughout this handbook— that must be included in any comprehensive, ecologically valid description of the pain experience (reviewed by Kucyi and Davis, 2015, 2016). The contents of spontaneous thoughts are also critical to consider clinically in acute and chronic pain. In this chapter, I review behavioral studies of ongoing fluctuations in attention to pain, studies of the neural basis of spontaneous mind-wandering away from pain, and the clinical implications of this research.

**Spontaneous Attentional Fluctuations and Pain**

Pain is an unpleasant sensory and emotional experience that can be measured only by self-report. Advances in functional neuroimaging allow prediction of self-reported acute pain states with greater than 90% accuracy under certain contexts (Wager et al., 2013, Woo et al., 2017). Importantly, these applications rely on comparison with self-report as the gold standard measure, and there is currently no validated technology that can objectively confirm or rule out the presence of pain (reviewed by Davis et al., 2015). Reliance on self-report, however, has several shortcomings and may be inadequate for fully capturing all aspects of the pain experience (reviewed by Wager and Atlas, 2013).

When assessing pain-attention interactions, one major issue is that being asked to rate pain inherently biases attention toward that pain. In daily life, attention may naturally wax and wane. Pain qualities cannot be reported on in a valid manner during moments when attention is on something other than pain. Indeed, people experiencing prolonged pain may not commonly be in states in which attention is fully engaged with features such as the exact current intensity or unpleasantness of pain.

One way to study how attention to pain naturally varies over time is to use the experience sampling approach. This method, in which people are probed at random intervals about their attention, is similar to that often used in the study of spontaneous thought (reviewed by Smallwood and Schooler, 2006). In a handful of studies, experience sampling has been used in patients with chronic pain, who were probed about their level of attention to pain (e.g. rating the statement "Right now, I am focusing on my pain") at random intervals during daily life (Roelofs et al., 2004, Viane et al., 2004, Peters and Crombez, 2007, Crombez et al., 2013). These studies confirm that attention varies naturally over time, and that some patients tend to attend away from pain more than others. However, it has not been fully determined whether attention away from pain is typically due to spontaneous thought or to other distractors (e.g. externally-driven events). Given that mind-wandering occurs frequently in everyday waking life (Kane et al., 2007), it is likely that spontaneous thoughts commonly drive attention away from pain in daily life, and that the content and frequency of these thoughts vary both within and across individuals.

While interactions of spontaneous thought with chronic pain remain to be characterized, significant advances have been made in the study of acute pain. In research that combined painful stimulation, experience sampling, and neuroimaging, 51 healthy adults were asked after several 20-second trials of painful (transcutaneous electrical) stimulation whether their attention had just been on pain or on something else (Kucyi et al., 2013). Although stimulus intensity was kept constant to evoke a predetermined pain intensity level, the degree of self-reported attention varied across trials. At the group-level, subjects reported on average that thoughts away from pain were mostly due to mind-wandering (i.e., thoughts completely unrelated to the stimulus or other features of the present sensory environment), but sometimes attention away from pain was due to distinct factors such as external distractions (e.g., hearing sounds). Importantly, subjects were tested with this experience sampling paradigm on 2 separate days, and the degree of self-reported attention away from pain was highly consistent between sessions, suggesting that people may have trait-like tendencies that predispose them to either attend to pain or instead become immersed in spontaneous thoughts.

Supporting this idea of trait-like tendencies, in a separate, demanding cognitive task (without experience sampling), individual differences in self-reported mind-wandering away from pain (as previously recorded with experience sampling) were predictive of the effect of pain on behavior. Specifically, reaction time showed greater slowing in the presence of pain in subjects who had reported more attention to pain (Kucyi et al., 2013). This behavioral link, together with brain activity measurements (described below), was critical to validating the self-reports obtained with experience sampling.

**Neural Basis of Mind-wandering Away from Pain**

The relationship between brain activity and spontaneous thought during painful stimulation must be interpreted in light of what is currently known about pain- and attention-related brain networks. Critically, nociceptive signals in the peripheral nervous system (the sensory response to harmful or potentially harmful stimuli) do not always result in pain. Pain arises from a specific pattern of dynamic brain activity that may sometimes get engaged, and may at other times not get engaged, by the same input stimulus (reviewed by Kucyi and Davis, 2015). When engaged, ascending spinal cord and brainstem pathways send signals to several regions within the cerebral cortex (including insula, somatosensory cortex, and cingulate cortex) (Apkarian et al., 2005) that are thought to reflect the characteristic sensory, cognitive-affective, and motivational aspects of pain perception (Melzack and Casey, 1968). When not engaged, a separate system known as the descending pain modulatory system (or antinociceptive system) is likely to be at play. This descending system, first characterized in animal models (reviewed by Basbaum and Fields, 1984) and now with supporting evidence from human functional neuroimaging studies (reviewed by Tracey and Mantyh, 2007), is thought to include specific areas within the cerebral cortex (including prefrontal subregions) that project to key brainstem nodes (periaqueductal gray [PAG] and rostroventral medulla) that inhibit incoming nociceptive input from the spinal cord.

Importantly, modern neuroimaging has revealed pain-related networks that may be preferentially involved in attentional aspects of pain. The salience network (Seeley et al., 2007), a bilateral system including the anterior insula and an anterior part of the temporoparietal junction, is activated with general, salient changes in the environment arising from any input modality, including pain (Downar et al., 2000, Downar et al., 2003), and some subregions of this network show greater within-network connectivity for right- compared to left-hemisphere homologous regions (Kucyi et al., 2012a, Kucyi et al., 2012b). Attentional modulation of pain through active distraction has been shown to decrease activity in areas within the salience network and to engage the descending pain modulatory system (reviewed by Seminowicz and Davis, 2007, Kucyi and Davis, 2015). Another system known as the default mode network (DMN), which generally deactivates when attention is directed to the external environment (Raichle et al., 2001), is typically deactivated during pain (Coghill et al., 1994, Loggia et al., 2012). A highly consistent effect in fMRI studies is that the DMN shows increased activation during self-reported mind-wandering (Fox et al., 2015), although recent studies have revealed a non-exclusive role in self-reported mind-wandering and potential importance to other aspects of attention (Crittenden et al., 2015, Kucyi et al., 2016a).

When healthy adults performed experience sampling during fMRI with noxious stimulation, several patterns of state-related activity were found in pain- and attention-related networks (Kucyi et al., 2013) (summarized in **Figure 1**). First, self-reported attention to pain was associated with greater activation within right hemisphere areas of the salience network (particularly anterior insula, dorsolateral prefrontal subregions, and temporoparietal junction) as well as regions implicated in a distinct "frontoparietal control network" (anterior intraparietal sulcus, dorsolateral prefrontal subregions). The engagement of salience network regions, particularly those in the right hemisphere, during sustained attention to pain is consistent with previous work (Downar et al., 2003) and with the notion the automatic, attention-demanding quality of pain is reflected in this network. While more speculative, the activity of regions in the frontoparietal control network could reflect a 'high jacking' via interactions with active regions of the salience network to impair the ability to shift away from pain.

A second finding was that self-reported attention away from pain was associated with lesser deactivation of the DMN, including medial prefrontal cortex (mPFC) and posterior cingulate cortex (PCC)/retrosplenial cortex, as well as areas within medial temporal lobe and dorsomedial prefrontal DMN subsystems. The subjects who reported that their attention away from pain was due to a high degree of mind-wandering (as opposed to external distraction) showed the greatest decrease of DMN deactivation. Finally, during self-reported attention away from pain, there was also increased functional connectivity (greater correlation of signals) between the PAG, a key node of the descending pain modulatory system, and core areas of the DMN (including mPFC, PCC, and retrosplenial cortex).

Further insights came from study of individually varying tendencies in relation to brain structural and functional connectivity (Kucyi et al., 2013). An analysis of diffusion MRI data revealed that individuals who reported more frequent attention away from pain were found to have higher fractional anisotropy [a measure potentially indicating stronger, or more intact, structural connectivity (Johansen-Berg and Rushworth, 2009)] in the pathway between the mPFC (within DMN) and PAG. Additionally, the spontaneous activity between these same two regions during ~9 minutes of wakeful rest was similarly related to individual differences. Subjects who reported more frequent attention away from pain had greater functional connectivity variability (possibly indicating more dynamic communication) between the mPFC and PAG (Kucyi et al., 2013). Spontaneous resting-state functional connectivity on the time-scale of minutes is well known to be largely reflective of intrinsic, individual-specific functional anatomy (reviewed by Buckner et al., 2013), and so these findings further support the idea of a trait-like nature of the tendency to mind-wander from pain (Kucyi et al., 2013). Taken together, the work suggests that brain activity underlying mind-wandering away from pain may share some similarities with active distraction from pain (e.g., decreased salience network activity), but that idiosyncrasies are also likely (e.g., decreased DMN deactivation and interaction with the descending pain modulatory system).

The study of the neural basis of mind-wandering and pain is in its infancy, and several caveats of the described neuroimaging findings should be considered. Perhaps most critically, to classify a neural process as truly antinociceptive, it is necessary to establish that the process underlies decreased pain. When measuring self-reported attention with experience sampling, pain ratings cannot be taken together with attention ratings. Possible interference with spontaneous attentional fluctuations may occur, and the validity of such ratings when attention is away from pain would be questionable. Thus, independent experiments, or a different paradigm, would be needed to confirm that mind-wandering away from pain involves antinociception (cf. Krakauer et al., 2017). Further experiments are also needed to determine the neural correlates of different types of spontaneous thoughts (e.g., past- versus future-related, positive versus negative emotional content, etc.). Another critical question for future research is whether altered communication occurs between the DMN and antinociceptive system during all mind-wandering (i.e., away from any sensory modality) or specifically during mind-wandering away from pain.

Although studies in non-human animals can help to provide insights when generating hypotheses, the nuanced self-reports needed to measure interactions between spontaneous thought and pain limit direct studies of the underlying physiological processes to humans— and thus, mainly to non-invasive neuroimaging modalities with limited spatial and temporal resolution. Invasive studies of structural connectivity in non-human primates provide support for plausible direct neuroanatomical pathways underlying the human findings that emphasize a role of connectivity between DMN areas and the PAG. Evidence indicates that areas within mPFC and within retrosplenial cortex have efferent connections to the PAG (An et al., 1998, Parvizi et al., 2006).

Further evidence is needed to demonstrate that the full antinociceptive system (beyond PAG; see model in **Figure 1**) is spontaneously engaged during mind-wandering away from pain. Key brainstem and spinal cord nuclei within the antinociceptive system are relatively small in size (sometimes below the scale of millimeters) and are difficult to study with fMRI due to technical limitations. Complicating matters, the PAG is comprised of several subregions with distinct connectivity and function (Linnman et al., 2012, Coulombe et al., 2016). Recently, advances with relatively high-resolution human spinal cord fMRI have allowed detection of the modulation of nociceptive signals in placebo analgesia (Eippert et al., 2009) and in active distraction from pain (Sprenger et al., 2012). The applications of spinal cord functional imaging and high-resolution brainstem imaging could be fruitful in the study of the antinociceptive system during paradigms involving measures of both pain and spontaneous thought.

Finally, the neurochemical basis, electrophysiological dynamics, and causal neural circuitry of mind-wandering away from pain remain unknown. The PAG is rich in opiate-containing neurons that mediate endogenous functions of the antinociceptive system (reviewed by Millan, 2002), and further studies could provide insight into whether fluctuations within the opioidergic system co-occur with spontaneous thought in the context of pain. Feedforward and feedback communication cannot be determined from fMRI changes, whereas frequency-specific signals from electrophysiological measurements could give mechanistic clues pointing toward the nature of dynamics within the salience network and DMN during mind-wandering away from pain (reviewed by Ploner et al., 2017). Interventions that involve perturbation of activity within these networks, for example with electrical brain stimulation, could be critical to establishing the causal roles of various structures in spontaneous levels of attention to pain. In summary, findings to date provide a platform for testable hypotheses regarding detailed neural mechanisms of interactions between pain and spontaneous thought.

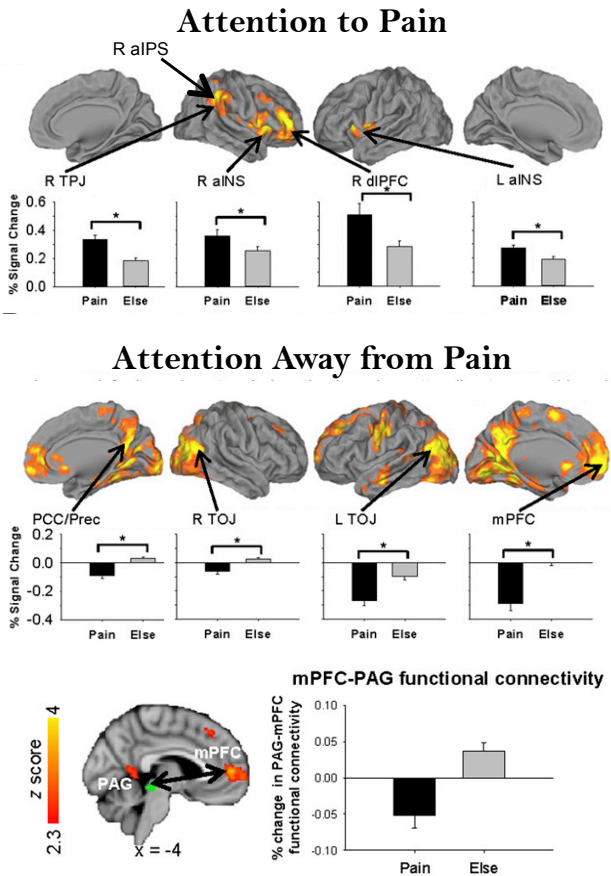

**Figure 1**. Brain dynamics of self-reported attention to and spontaneous mind-wandering from pain. (Top) Brain regions within salience and frontoparietal networks show increased activation during attention to compared to away from pain. During attention away compared to toward pain, brain regions in the default mode network show decreased deactivation (Middle) and the periaqueductal gray shows increased functional connectivity with the medial prefrontal node of the default mode network. aINS = anterior insula, aIPS = anterior intraparietal sulcus; mPFC = medial prefrontal cortex, dmPFC = dorsomedial prefrontal cortex, dlPFC = dorsolateral prefrontal cortex, MTL = medial temporal lobe, mPFC = medial prefrontal cortex, PAG = periaqueductal gray, PCC = posterior cingulate cortex, Prec = precuneus; TOJ = temporo-

occipital junction; TPJ = temporoparietal junction. Adapted with permission from (Kucyi et al., 2013).

**Clinical Implications**

Although not often explicitly considered within clinical contexts, a close look at the pain field reveals that clinicians and researchers widely recognize the potentially important role of spontaneous thought in coping with chronic pain. Commonly used clinical scales include inquiries about the temporal fluctuations of pain [e.g., the McGill Pain Questionnaire (Melzack, 1975) and painDETECT (Freynhagen et al., 2006)], including some scales that directly inquire about how pain tends to intrinsically capture attention. The Pain Catastrophizing Scale (Sullivan et al., 1995) includes a subscale on rumination about pain, defined as perseverative negative thinking about pain and its possible causes or consequences [i.e., a lack of spontaneity in thought (Christoff et al., 2016); see also the chapter by DuPre and Spreng in this volume]. The Pain Vigilance and Awareness Questionnaire (McCracken, 1997) and Experience of Cognitive Intrusion of Pain scale (Attridge et al., 2015) include probes about the tendency to attend to or mind-wander away from pain. Spontaneous thoughts are predominantly accompanied by positive or neutral affect (Fox et al., 2014) and thus could be protective against negative emotions commonly associated with rumination about and excessive attention to pain.

Individual variability in rumination about chronic pain has been studied both behaviorally and at the neural level. Studies of chronic pain populations have revealed that patients with a greater tendency to ruminate about pain tend to experience a greater level of pain and have poorer clinical outcomes (Sullivan et al., 2002, Van Damme et al., 2002, Buenaver et al., 2012). In chronic pain patients with temporomandibular disorder, enhanced resting-state functional connectivity within the DMN was associated with greater rumination about pain (Kucyi et al., 2014). The same study also found a positive correlation between rumination and functional connectivity between the mPFC and PAG areas (among others) (Kucyi et al., 2014). This could suggest a compensatory mechanism, given that mPFC-PAG functional connectivity increases during mind-wandering away from pain (Kucyi et al., 2013). While intriguing, such findings should be replicated and further extended in independent cohorts and in other chronic pain populations.

A theme in this chapter has been that the tendency to attend to spontaneous thoughts in the presence of nociceptive input may be a trait-like quality. However, an important, unanswered question concerns whether trainable cognitive states could allow patients to overcome excessive attention to pain in tandem with reorganization of brain structure and function. There has been considerable recent progress in development of and research on mindfulness meditation-based training for chronic pain, in which patients are encouraged to attend to and accept sensory (but not affective) aspects of their pain from a non-evaluative standpoint (reviewed by Zeidan and Vago, 2016). While mindfulness is thus proposed to work therapeutically via a specific form of *enhanced* attention to pain, there is currently no comparable, established behavioral treatment that specifically focuses on *reducing* attention to pain (e.g., increasing spontaneous thoughts away from pain and/or reducing rumination about pain). If an intervention could reliably increase levels of mind-wandering away from pain, a testable hypothesis would be that such a therapy is effective in patients who may not benefit from mindfulness.

The practice of cognitive-behavioral therapy (CBT) for pain, a structured psychotherapeutic approach, involves training patients to actively control their pain and the associated negative affect (Turk et al., 1983, Thorn, 2004). A goal of this approach is to get learned coping strategies to trickle into everyday spontaneous coping. Potential effects on the tendency to mind-wander from pain could be considered in future studies as a clinical outcome measure. Interestingly, cognitive-behavioral training has been shown to alter DMN resting connectivity as well as DMN deactivation during acute pain (Kucyi et al., 2016b). Also, in chronic pain, CBT led to increased resting connectivity between DMN areas and the PAG (among additional effects in other networks) (Shpaner et al., 2014). These neural changes in brain systems relevant to spontaneous pain-attention interactions suggest that CBT could influence the tendency to mind-wander, but independent studies directly testing this hypothesis are needed.

Uncovering the brain mechanisms of spontaneous thought in the context of chronic pain could inform the development of neurorehabilitation strategies. Based on available data in acute pain and preliminary findings in chronic pain, pathways between the DMN and descending pain modulatory system (e.g., between mPFC and PAG) could represent potential neuromodulatory targets for alleviating excessive attention to and rumination about pain. However, the causal roles of these pathways in spontaneous pain-attention interactions, and the potential importance of broader networks associated with these pathways, have not yet been clarified. It also remains to be seen whether the brain dynamics of mind-wandering away from acute pain are representative of what occurs in chronic pain. Thus, any development of a relevant, potentially effective therapy must be informed by detailed future studies with the specific focus on qualities of spontaneous thought, and associated brain mechanisms, in patients with chronic pain.